\documentstyle[preprint,eqsecnum,aps]{revtex}

\begin{document}

\draft

\title{The Weyl anomaly and the nature of the background geometry}

\author{ Israel Quiros\thanks{israel@uclv.etecsa.cu}}
\address{ Departamento de Fisica. Universidad Central de Las Villas. Santa Clara. CP: 54830 Villa Clara. Cuba }

\date{\today}

\maketitle

\begin{abstract}

The Weyl anomaly problem is treated within a purely geometrical context. Arguments are given that hint at a possible classical origin of the conformal anomaly in the Riemannian nature of the background geometry where the matter fields play out their dynamics. Some considerations allowing for a possible resolution of the Weyl anomaly problem are briefly outlined. Following the spirit of the standard model of the fundamental interactions, it is argued that the Weyl anomaly should be a consequence of the breaking of the gauge symmetry at some stage during the evolution of the universe.

\end{abstract}


The usefulness of Riemann geometry in relativity has been questioned both from the axiomatic and the observational points of view. In Ref.\cite{eps}, for instance, the authors tried to give an axiomatical foundation for the geometrical nature of spacetime by means of ideal operations with elementary clocks and rods. Following this procedure one finds that the Riemannian structure of spacetime is not well-founded. Instead a Weylian structure is more appropriate\cite{vp}. From the observational standpoint one has that the measuring procedure must involve, besides the metric tensor $g_{ab}(x)$, a scalar function $\omega(x)$ allowing for the conformal invariance of null cones that is one of the most important observational aspects of the background geometry. This leads a conformally-Riemannian structure to be involved rather than a Riemannian one\cite{ja}.

The appearance of two geometries in a single theory of gravity is familiar\cite{bekenstein}. Yet in the early 60-ies Brans and Dicke\cite{bdk} raised questions about the usefulness of Riemann geometry in relativity due to the arbitrariness in the metric tensor which results from the indefiniteness in the choice of the units of measure. As in Brans-Dicke-type theory so in Dirac's theory of variable gravitational constant\cite{dirac}, string theory\cite{gsw} and many others, two conformally related geometries appear. One of them usually describes gravitation while the other defines the geometry in which matter dynamics takes place. Other alternatives for the geometry where the matter plays out its dynamics have been studied. In Ref.\cite{bekenstein}, for instance, Bekenstein supposed the physical metric to be Finslerian.

However, the most promising alternatives to a Riemann structure of spacetime are the conformally-Riemannian (Weylian) configurations. Besides the former considerations, these alternatives are promising for two reasons. First, Weyl structures possess an additional degree of freedom to make gauge transformations\cite{adler} and this makes it interesting for dealing with the gauge principle in field theory\cite{scipioni}. Second, there are some effects that take place in Weyl spaces, that mimic classically some purely quantum effects. In Ref.\cite{novello}, for instance, a classical analogue of a quantum tunneling process in a spacetime of Weyl-integrable configuration was studied. In Ref.\cite{dzhunu} the authors demonstrated that there is some kind of correlation between objects in classical gravity on Weyl manifolds and in quantum non Abelian field theory. However, the first (and, may be, the soundest) example of such quantum behavior of classical Weyl structures was given by London\cite{adler,london}. He demonstrated that certain intrinsic, quantum mechanical rules can be associated with Weyl geometry with an imaginary fine structure constant. Other hints at connection between Weyl (and similar) structures and quantum behavior can be found in Ref.\cite{rankin}.

A spacetime of Weylian structure is an affine manifold specified by a metric tensor $g_{ab}(x)$ and a gauge vector $\omega_a (x)$ which enter the definition of the manifold Weyl affine connection\cite{adler,weyl}

$$
\Gamma^a_{bc}=\{\;^a_{bc}\}-\frac{1}{2}(\omega_c\delta^a_b+\omega_b\delta^a_c-g_{bc}\omega^a),
\eqno{(1)}
$$
where $\{\;^a_{bc}\}=\frac{1}{2}g^{an}(g_{bn,c}+g_{cn,b}-g_{bc,n})$ are the Christoffel symbols of the metric. The rule of parallel transport of a given vector in Weyl geometry requires a non vanishing covariant derivative of the metric tensor\cite{novello}

$$
g_{ab;c}=\omega_c g_{ab},
\eqno{(2)}
$$
where the semicolon denotes covariant differentiation in a general affine sense, i.e., through the Weyl connection Eq.(1).

As already remarked, besides the manifold motions group of Riemann structures, Weyl geometries admit internal (gauge) transformations

$$
g_{ab}\rightarrow\Omega^2(x) g_{ab},
\eqno{(3)}
$$
and

$$
\omega_a(x)\rightarrow\omega_a(x)+2\Omega^{-1}\Omega_{,a}.
\eqno{(4)}
$$

The conformal transformation Eq.(3) is also acknowledged as a Weyl rescaling of the metric tensor. It is very encouraging that classical massless field systems (matter with a trace-free stress-energy tensor) that are in interaction with gravity, display conformal invariance under Weyl rescalings of the kind Eq.(3). In effect, usually, the stress-energy tensor of matter fields that are minimally coupled to the spacetime metric fulfil the following dynamic ('conservation') equation:

$$
T^{an}_{\;\;\;\;\|n}=0,
\eqno{(5)}
$$
where the double-bar means covariant derivative in a Riemannian sense, i.e., through the Christoffel symbols of the spacetime metric $\{\;^a_{bc}\}$. Under the Weyl rescaling Eq.(3) the condition Eq.(5) is transformed into the following condition:

$$
T^{an}_{\;\;\;\;\|n}+2\Omega^{-1}\Omega^{,a} T^n_n=0,
\eqno{(6)}
$$
where now the covariant (Riemann) derivatives are given in terms of the rescaled (conformal) metric. For matter fields with a trace-free stress-energy tensor ($T^n_n=0$), it is easily verified from Eq.(6) that the dynamic equation (5) is invariant under the Weyl rescaling Eq.(3). It is the classical result I have already remarked.

However, this invariance no longer survives in the quantum theory\cite{capper}. This situation is acknowledged as the 'Weyl anomaly' or 'trace anomaly' and an amazing number of papers has been devoted to this subject. Weyl anomalies have found a variety of applications in black hole (wormhole) physics, cosmology (inflation in the early universe, the vanishing of the cosmological constant in the present era, particle production), supergravity, and in superstring theory. For a readable review on this subject we recommend reading Ref.\cite{duff}.

The aim of the present paper is, precisely, to address the 'Weyl-anomaly' problem, being quantum in nature, at a purely geometrical (classical) context. I shall hint at a possible origin of this 'anomaly' in the geometrical structure of the background manifold where massless matter fields play out their dynamics. Some considerations allowing for a possible resolution of the 'Weyl anomaly' problem will be briefly outlined.

The fact I want to remark here is that in usual field theory coupled to gravity the background geometry is taken to be of Riemannian nature. The basic requirement of Riemann geometry is the vanishing of the covariant derivatives of the metric tensor at each point:

$$
g_{ab;c}=0.
\eqno{(7)}
$$

If this condition is fulfilled, the manifold symmetric affine connections $\Gamma^a_{bc}$ become identical to the Christoffel symbols of the Riemann metric\cite{adler} and $g_{ab;c}=0\Rightarrow g_{ab\|c}=0$. The following step in this line of reasoning is to realize that under the Weyl rescaling Eq.(3), the requirement Eq.(7) is transformed into the following non-Riemannian requirement:

$$
g_{ab;c}+2\Omega^{-1}\Omega_{,c} g_{ab}=0.
\eqno{(8)}
$$

This requirement means that the units of measure of the conformally-Riemannian geometry should have point-dependent length. In other words, the Riemannian nature of the background geometry is not invariant under the Weyl rescaling Eq.(3).

Up to this moment in the discussion, the main conclusion to be drawn is that, irrespective of the fact that, for massless matter fields, the classical 'conservation' equation (5) is conformally invariant, the nature of the background geometry where these matter fields play out their dynamics is not preserved by the conformal transformations of the metric Eq.(3). Otherwise, there is an obvious contradiction. On the one hand the dynamical (conservation) equation of massless fields (Eq.(5)) is gauge invariant, while, on the other hand, the nature of the background geometry where this equation is to be geometrically interpreted is not gauge invariant. This result hints at a possible geometrical origin of the Weyl anomaly. In effect, the fact that under the Weyl rescalings Eq.(3), the dynamical equation (5) has different invariance properties for massless fields and for matter fields with non zero mass, means that it is a less fundamental requirement than the Riemann requirement Eq.(7) that is not conformally invariant. We recall that all fields (both massless and with non zero mass) play out their dynamics on a background spacetime of Riemannian configuration. Hence, the origin of the Weyl anomaly should be in the (classical) Riemannian nature of the background geometry rather than (only) in the quantum aspects of the dynamics of matter fields coupled to gravity.

If the above considerations are correct, and the origin of the Weyl anomaly is in the Riemannian structure of the background geometry, one can wonder whether, by addressing the play out of the matter dynamics at other possible (non Riemannian) geometrical backgrounds, one can avoid the ocurrence of this anomaly. In this sense, and in the spirit of the considerations given in the introductory part of this paper, I shall ask for a possible resolution of the Weyl anomaly problem by approaching a geometrical background of Weyl nature. This choice of the nature of the background geometry is even more justified since, as properly remarked in Ref.\cite{feoli}, in the spirit of the electroweak model, Weyl geometry should be considered a more fundamental geometry than Riemannian one. In effect, Riemann geometry represents a state of broken conformal invariance.

When one approaches a background geometry of Weyl configuration, one has to realize that the basic geometrical requirement is that given in Eq.(2). This requirement is invariant under the gauge transformations Eq.(3) and (4), meaning that Weyl geometry is a conformally invariant configuration. If one chooses the matter dynamics to be driven by the conservation equation (5), one arrives again at a kind of 'conformal anomaly'. In effect, in this case the non zero mass matter fields are not conformally (gauge) invariant while the background (Weyl) geometry where they play out their dynamics is invariant under the gauge transformations Eq.(3) and (4). Therefore, if one looks for a possible resolution of the Weyl anomaly problem by approaching a bacground geometry of Weyl configuration one should, at the same time, to look for gauge invariant 'conservation' equations. Working in this direction one finds that the equation

$$
T^{an}_{\;\;\;\;\|n}=-\frac{1}{2}\omega^a T^n_n,
\eqno{(9)}
$$
or, equivalently

$$
T^{an}_{;n}=-3\omega_n T^{na},
\eqno{(10)}
$$
is invariant under the gauge transformations Eq.(3) and (4). An equation of the kind Eq.(9) (or Eq.(10)) is allowed in theories of gravity where the matter fields are non-minimally coupled to the scalar factor $\omega=\int dx^n\omega_n(x)$, through the Lagrangian ${\cal L}_{nm}=\sqrt{-g}e^{2\omega}L_{matter}$, where $L_{matter}$ is the Lagrangian density for the matter fields. In other words, in theories of gravity where the matter dynamics is driven by a Lagrangian of the kind ${\cal L}_{nm}$ (the dynamic equation (9) or (10) takes place) and where the dynamical behavior of the matter fields is geometrically interpreted within the context of Weyl geometry (the basic requirement is given in Eq.(2)), there is no place for the Weyl anomaly. Therefore, following the spirit of the standard (electroweak) model, theories with non-minimal coupling of the matter fields to the metric are more fundamental than theories with minimal coupling since, the latter theories are allowed only after a state with broken gauge symmetry is reached. Following this line of reasoning one is left to the conclusion that the Weyl anomaly is a result of the gauge symmetry breaking taking place at a given stage during the evolution of the universe.

It is honest noting, however, that approaching background geometries of Weyl configuration is problematic in one aspect. In effect, according to the definition of a Weyl space, variations of the units of measure are controled by the gauge vector $\omega_a(x)$. In particular, if $l=g_{nm}V^n V^m$ is the length of a given vector $V^a(x)$, in the course of an infinitesimal parallel transport $dx^a$, this length varies according to\cite{adler,weyl}

$$
dl=dx^n\omega_n(x)\;l.
\eqno{(11)}
$$

For closed paths a synchronization loss, additional to the usual loss of synchronization due to gravitational effects, would desagree with well-known observations\cite{vp}. It is due to the non-integrability of length according to Eq.(11). To overcome this objection, one has to impose coincidence of the units of measure regardless of the particular closed path chosen, implying that $\oint\frac{dl}{l}=0$. This last condition is fulfilled if $\omega_{a,b}-\omega_{b,a}=0$, that is, $\omega_a(x)=\omega(x)_{,a}$. In other words, if the gauge vector $\omega_a$ can be written as a gradient of some scalar function $\omega(x)$, then the length of the given vector is integrable along closed path and no additional synchronization loss occurs. Weyl geometries for which this condition is fulfilled are acknowledged as integrable-Weyl geometries\cite{novello}. Consequently, the next step in the chain of considerations given in this paper is to consider only integrable-Weyl background geometries. The basic requierement for these geometries is the following:

$$
g_{ab;c}=\omega_{,c} g_{ab},
\eqno{(12)}
$$
consequently, the dynamical 'conservation' equation for matter fields is

$$
T^{an}_{\;\;\;\;\|n}=-\frac{1}{2} g^{am}\omega_{,m} T^n_n,
\eqno{(13)}
$$
or, equivalently

$$
T^{an}_{;n}=-3\omega_{,n} T^{na}.
\eqno{(14)}
$$

Both equations (12) and (13) (or (14)) are invariant in respect to the following gauge transformations

$$
g_{ab}\rightarrow\Omega^2(x) g_{ab},
$$
$$
\omega(x)\rightarrow\omega(x)+\ln\Omega^2(x).
\eqno{(15)}
$$

This way, once again, the 'conformal anomaly' problem is overcome. By this time, however, matter fields play out their dynamics on background geometries of integrable-Weyl nature, so there are no observational objections respecting the additional loss of synchronization that is inherent to Weyl geometries in general.

Finally, I shall to remark that, following the spirit of the standard model of the fundamental interactions, one should regard integrable-Weyl structures as more fundamental than Riemannain ones. In the same way, non-minimal coupling of matter fields to the scalar field $\omega(x)$ through ${\cal L}_{nm}$ is more fundamental than minimal coupling. Therefore, as noted above in this paper, the Weyl anomaly should be a consequence of the breaking of the gauge symmetry at some stage in the evolution of the universe. More discussion on this subject will be given in future works.

I acknowledge useful converstaions with my colleagues Rolando Cardenas and Rolando Bonal and MES of Cuba by financial support of this research.

\end{document}